\documentclass[twocolumn,prb,showpacs,aps,superscriptaddress]{revtex4}

\usepackage{amsmath}
\usepackage{amsbsy}
\usepackage{amsthm}
\usepackage{amssymb}
\usepackage{latexsym}

\usepackage[dvips]{color,graphicx}

\begin{document}
\title{Supersolids in confined fermions on one-dimensional optical lattices}
\author{F. Karim Pour}
\affiliation{Institut f\"ur Theoretische Physik III,
Universit\"at Stuttgart, Pfaffenwaldring 57, D-70550 Stuttgart, Germany.}
\author{M. Rigol}
\affiliation{Department of Physics and Astronomy,
University of Southern California,
Los Angeles, California 90089, USA}
\author{S.\ Wessel}
\affiliation{Institut f\"ur Theoretische Physik III,
Universit\"at Stuttgart, Pfaffenwaldring 57, D-70550 Stuttgart, Germany.}
\author{A. Muramatsu}
\affiliation{Institut f\"ur Theoretische Physik III,
Universit\"at Stuttgart, Pfaffenwaldring 57, D-70550 Stuttgart, Germany.}

\begin{abstract}
Using quantum Monte Carlo simulations, we show that density-density 
and pairing correlation functions of the one-dimensional attractive 
fermionic Hubbard model in a harmonic confinement potential are 
characterized by the anomalous dimension $K_\rho$ of a corresponding 
periodic system, and hence display quantum critical behavior. The 
corresponding fluctuations render the SU(2) symmetry breaking by the 
confining potential irrelevant, leading to structure form factors for 
both correlation functions that scale with the same exponent upon 
increasing the system size, thus giving rise to a (quasi)supersolid. 
\end{abstract}
\maketitle

The recent experimental observation\cite{kim04a} of nonclassical rotational 
inertia in solid $^4$He revived the interest in the possible coexistence 
of superfluidity and crystalline order, a phase named a supersolid (SS), proposed long ago,\cite{andreev69} in spite of arguments against 
it.\cite{penrose56} Such arguments seem to be validated by quantum 
Monte Carlo (QMC) simulations\cite{ceperley04} and theoretical work,\cite{prokofev05} showing that a defect-free crystal of $^4$He 
does not show off-diagonal long-range order (ODLRO). Further numerical 
simulations suggest that ODLRO may arise from domain walls\cite{burovski05} 
or in a metastable state,\cite{boninsegni06} such that the existence 
of a pure supersolid phase in $^4$He is still under debate.\cite{kim06}
 
Alternative paths toward the realization of a SS were recently suggested 
for quantum gases on optical lattices, e.g., for bosons with 
dipolar interactions,\cite{goral02} Bose-Fermi mixtures,\cite{buechler03}
Bose-Bose mixtures,\cite{mathey06} frustrated lattice 
geometries,\cite{frustbose} or extended interactions\cite{sengupta05} 
by loading the atoms in higher bands.\cite{scarola05} Most of the 
proposals, however, are based on mean-field approximations that tend 
to be unstable with respect to phase separation, as shown by QMC 
simulations.\cite{sengupta05} Also the presence of a confining potential 
is in general neglected, discarding the coexistence of 
phases\cite{batrouni02,rigol03} that may obscure the 
experimental determination of a SS.\cite{scarola06}

Another system where a SS phase is well known to exist\cite{micnas90} 
is the fermionic Hubbard model with an attractive contact interaction 
[the first two terms in Eq.\ (\ref{Hamiltonian}) below]. In contrast 
to the mentioned bosonic systems, where vacancies or defects lead to a 
SS, here diagonal and ODLRO result from an SU(2) symmetry, 
relating density and pairing amplitudes at density $n=1$.\cite{micnas90} 
While the experimental realization of such a model is at least debatable 
in solid state systems, fermions with attractive contact interaction 
were recently achieved with quantum gases by means of Feshbach resonances, leading to pairing and superfluidity.\cite{regal04,bartenstein04,zwierlein04} 
Even the observation of superfluidity with fermions in an optical lattice 
for a dense system ($n \sim 1$) was announced recently,\cite{chin06} 
bringing very close the realization of an SU(2)-symmetric SS.

Still, the presence of a confining potential could inhibit the formation 
of the SS, since it explicitly breaks the above-mentioned SU(2) symmetry, 
as discussed in detail below. We show, however, based on QMC simulations 
of the one-dimensional (1D) attractive fermionic Hubbard model 
in a harmonic potential, (i) that density-density (DD) and pairing 
correlation functions are characterized by the anomalous 
dimension\cite{giamarchi04} $K_\rho$ of the corresponding periodic
model, and (ii) that in spite of the confinement, conditions can be 
reached where the structure form factors of both correlation functions 
diverge with the same power as the system tends toward the thermodynamic 
limit. Hence, a proper treatment of long-ranged quantum fluctuations 
shows that the SU(2) symmetry breaking becomes irrelevant, 
thus giving rise to a SS state.

In the following, we consider the Hamiltonian
\begin{eqnarray}
\label{Hamiltonian}
H & = & -t \sum_{j,\sigma} 
( c^\dagger_{j\sigma} c^{}_{j+1 \sigma} + h.c.)
- U \sum_j n_{j \uparrow} n_{j \downarrow} 
\nonumber \\ & &
+ V \sum_{j=0}^N \left(x_j - \frac{Na}{2} \right)^2 n_{j},
\end{eqnarray}
where $c^\dagger_{j\sigma}$ and $c^{}_{j\sigma}$ are creation and 
annihilation operators, respectively, for fermions on site $j$ 
(position $x_j=ja$, where $a$ is the lattice constant) with spin 
$\sigma = \uparrow, \downarrow$. The local density is 
$n_j=n_{j \uparrow} + n_{j \downarrow}$, where $n_{j \sigma} = c^\dagger_{j\sigma} c^{}_{j \sigma}$.
The contact interaction is attractive ($U>0$) and the last term 
models the potential of the confining trap with $N$ (even) sites. 
The QMC simulations were performed using a projector 
algorithm\cite{sugiyama86,sorella89,loh92} ($T=0$) with a projector 
parameter $\theta \simeq 30/t$ that suffices to reach well-converged 
values of all observables discussed here. For time slices below 
$\Delta \tau = 0.1/t$, the corresponding systematic error has no 
effect on the results. We use in general $\Delta \tau = 0.08/t$. 

It is well known\cite{micnas90} that a particle-hole transformation 
$d^\dagger_{j\uparrow} = \left(-1\right)^j c_{j\uparrow}$, 
$d^\dagger_{j\downarrow} = c^\dagger_{j\downarrow}$ 
applied to the Hamiltonian (\ref{Hamiltonian}) leads to a repulsive 
Hubbard model in the presence of a spatially varying magnetic 
field $h_j = 2V(x_j - Na/2)^2$. If
$\langle n_{j\uparrow}\rangle = \langle n_{j\downarrow}\rangle$, then 
$n^d_j \equiv \langle d^\dagger_{j\downarrow}d^{}_{j\downarrow} 
\rangle+ \langle d^\dagger_{j\uparrow}d^{}_{j\uparrow}\rangle =1$,
i.e., the new system is half filled. If, furthermore, $h_j=0$, i.e., 
for $V=0$, a Mott insulator with dominating antiferromagnetic 
correlations results for the repulsive system.\cite{giamarchi04}
In terms of the operators of Eq.~(1), this implies that for $V=0$ and 
$n=1$ the DD correlation function $N_{j\ell} \equiv \langle n_j 
n_\ell\rangle -\langle n_j\rangle \langle n_\ell\rangle $
and the pairing correlation function 
$P_{j \ell} \equiv \langle \Delta^{}_j \Delta^\dagger_\ell\rangle $, 
with $\Delta_j \equiv c_{j\uparrow} c_{j\downarrow}$, obey 
\begin{equation}
\label{NandP}
N_{j\ell} = 2 (-1)^{\mid j-\ell\mid} P_{j\ell},
\end{equation} 
and will be the dominating correlation functions, with the same 
power-law decay. This identifies the state as a SS. However, the 
presence of a confining potential breaks this SU(2) symmetry, so 
that for $V \neq 0$ the SS is expected to be destroyed. In the 
following we show that, due to quantum critical fluctuations, 
even for $V \neq 0$ the SS is nevertheless recovered.

\begin{figure}[t]
\includegraphics[width=.3\textwidth,angle=-90]{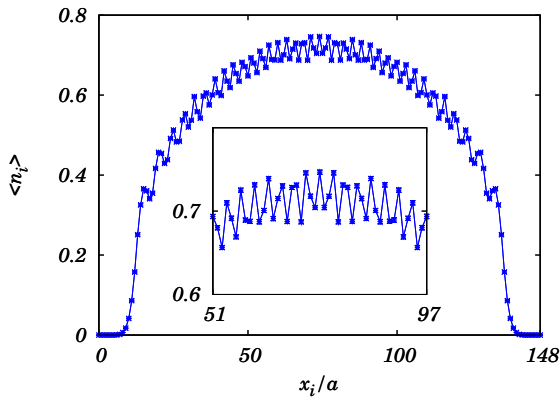}
\caption{Density $\langle n_i\rangle $ vs position $x_i$ in units of
the lattice constant $a$ for $N_f=74$ fermions in a harmonic confining 
potential $Va^2/t=1.826 \times 10^{-4}$, with $U/t=4$. The inset shows 
a detailed view of the density oscillations.}
\label{Density}
\end{figure}

A first important question is whether, in the presence of a confining 
potential, which produces a spatial dependence of the density, the 
correlation functions will each exhibit a decay with a single power. 
In fact, this is no longer the case if one relies on the local density 
approximation (LDA).\cite{butts97} However, we find that the LDA is 
not appropriate to describe the physics of the model in Eq.~(1).
In Fig.\ \ref{Density} we show the density profile for $N_f=74$ 
fermions, $U/t=4$, and a characteristic density $\tilde \rho =1$. 
As shown before,\cite{rigol04a} the characteristic density defined 
as $\tilde \rho = N_f a(V/t)^{1/2}$ relates systems with different 
sizes, number of particles, and confining harmonic potentials in the 
same way as the particle density does for periodic systems of different 
sizes. Figure \ref{Density} shows that the presence of the confining 
potential produces density oscillations on a short distance scale 
($\sim a$), which are absent in periodic systems. These oscillations 
preclude the use of the LDA as a sensible approximation, since it 
would require a slowly varying density.

\begin{figure}[t]
\includegraphics[width=.43\textwidth]{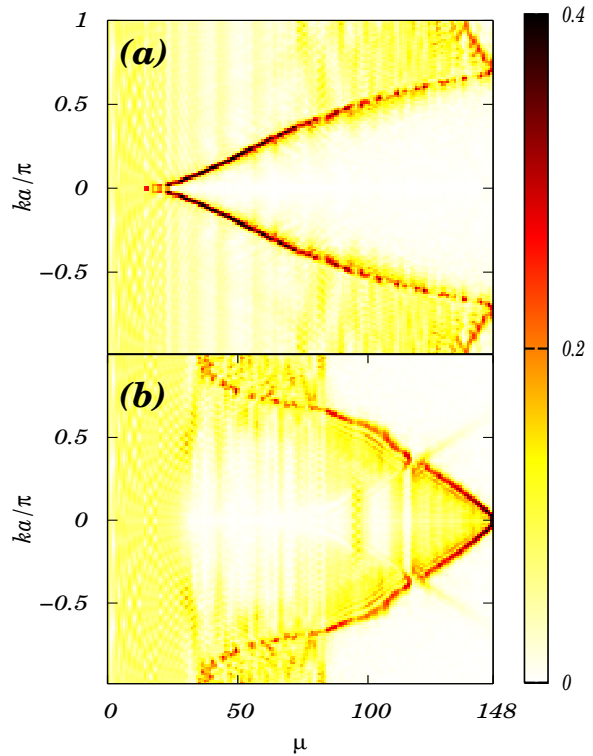}
\caption{(Color online) Level plot of the moduli of the Fourier 
transforms of $\phi_j^\mu$ (a) and $\varphi_j^\mu$ (b) vs $\mu$ 
[see Eqs.\ (\ref{DiagonalizationOfNiij}) and (\ref{DiagonalizationOfPiij})].}
\label{Eigenvectors}
\end{figure}

We study next the DD and pairing correlations using QMC simulations. 
Since in the presence of quasi-long-range order a divergence
is expected for a particular wave vector $k$ in a periodic system, 
we examine the eigenvalue equation for DD correlations,
\begin{equation}
\label{DiagonalizationOfNiij}
\sum_\ell N_{j\ell} \, \phi^\mu_\ell = N^\mu \phi^\mu_j,
\end{equation}
which reduces to a Fourier transformation for a periodic system,
i.e., the eigenvectors $\phi^\mu_j$ are plane waves, each mode 
characterized by a particular $k$ vector. Interestingly, we find 
that even inside a harmonic trap most of the modes $\phi^\mu_j$ 
can still be assigned each to a characteristic wave vector.
This can be seen by considering the moduli of the Fourier transform 
[$\sim \sum_j \phi^\mu_j \exp \left(i k x_j\right)]$ of each $\phi^\mu_j$.
As shown in Fig.\ \ref{Eigenvectors}(a), most of the modes $\mu$, with 
$N^\mu$ in ascending order, have a dominant component each of which
can be related to a wave vector $k$ (actually, both $\pm k$
due to reflection symmetry). We checked that regions where a clear 
identification of a dominant $k$ vector is not possible at all correspond 
to the lowest eigenvalues that arise from the vanishing density at the 
edges of the fermionic cloud in Fig.\ \ref{Density}. 
Results of a similar analysis for the eigenvectors $\varphi^\mu_\ell$ 
of the pairing correlations $P_{j\ell}$, 
\begin{equation}
\label{DiagonalizationOfPiij}
\sum_\ell P_{j\ell} \, \varphi^\mu_\ell = P^\mu \varphi^\mu_j,
\end{equation}
are shown in Fig.\ \ref{Eigenvectors}(b). We find that the highest 
eigenvalues of the pairing correlation function have the largest 
weight around $k=0$, i.e., pairs are formed by fermions with 
opposite momenta. This unbiased result helps to understand why even in
trapped systems, where translation invariance is broken 
and momentum is not a ``good'' quantum number, pairing correlations 
have been observed predominantly between atoms with opposite 
momenta.\cite{greiner05}

\begin{figure}[t]
\includegraphics[width=.4\textwidth]{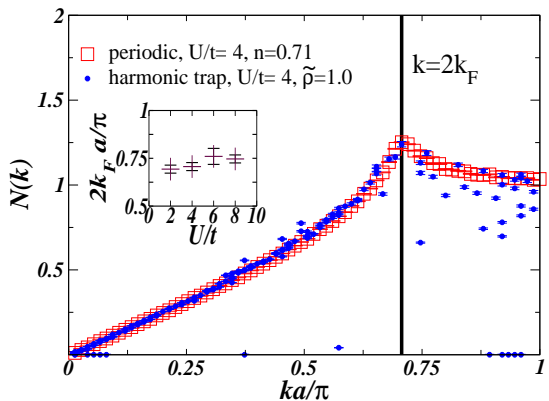}
\caption{(Color online) $N(k)$ for the confined ($\bullet$) case with 
the same parameters as in Fig.\ \ref{Density}. Superimposed are the 
results for a periodic system ($\Box$) with $2 k_F$ given by the black 
vertical line. The resulting density is $n = 0.71$. Inset: position of 
the $2 k_F$ peak as a function of $U$ in the harmonic trap for 
$\tilde \rho = 1$.}
\label{2kF}
\end{figure}
Since each mode $\mu$ can be assigned rather clearly to a wave vector 
$k$, we can map each eigenvalue $N^\mu$ to $N(k)$ as in a periodic 
system. Figure \ref{2kF} shows the result of such an assignment 
($\bullet$), according to the salient features shown in 
Fig.\ \ref{Eigenvectors}(a). We superimposed in Fig.\ \ref{2kF} the 
results for a periodic system ($\Box$), for which the density was fixed 
by identifying the wave vector corresponding to the maximum in $N(k)$ 
of the confined system with $2 k_F$ of the periodic system (black vertical 
line), where $k_F$ is the Fermi momentum, establishing in this way a link 
between a trapped system with a characteristic density $\tilde \rho$ and 
a periodic system of density $n = 2 k_F a/\pi$. Such an identification 
is based on the observation that, in a 1D fermionic system with attractive
interactions, $2 k_F$ oscillations in the charge channel are dominating. 
The inset shows the position of the peak as a function of $U$, which for 
the range $2 \leq U/t \leq 8$ does not change appreciably.
Hence, the identification is possible for a wide range of parameters and 
does not depend on any particular fine tuning.

While the identification of $N(k)$ in the periodic and the confined 
systems deteriorates at wave vectors beyond $2 k_F$ (with a number of 
points at zero or close to it due to the regions where the density 
vanishes), it is remarkably good close to $k=0$. For a periodic system 
the DD correlation function obeys
$N(k) \rightarrow K_\rho |k| a /\pi$ for $k \rightarrow 0$.\cite{giamarchi04}
Therefore, the anomalous dimension $K_\rho$ that determines the power-law 
decay of the $2 k_F$ oscillations that dominate at large distances
can be directly obtained from the slope of $N(k)$ for $k \rightarrow 0$. 
Figure \ref{2kF} shows that the mapping proposed here leads to the 
identification of $K_\rho$ also for the confined system.
\begin{figure}[t]
\includegraphics[width=.4\textwidth]{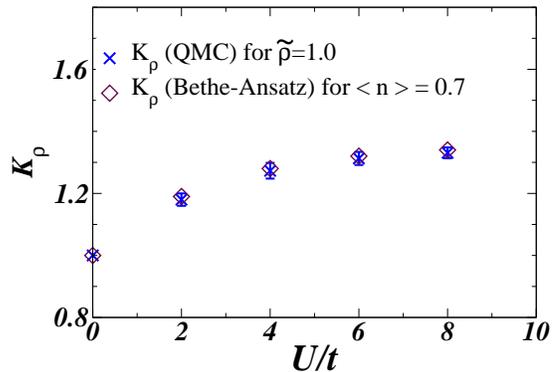}
\caption{(Color online) Anomalous dimension $K_\rho$ extracted from the 
slope of $N(k)$ for the confined system ($\times$) vs $U$ for 
$\tilde \rho = 1.0$ and $\langle n \rangle = 0.71$. 
Superimposed are the results for a periodic system obtained from Bethe 
ansatz ($\diamond$) for $\langle n \rangle = 0.7$ 
(Ref.\ \onlinecite{giamarchi04}).}
\label{Krho}
\end{figure}
In fact, such an identification is possible in general, as shown in 
Fig.\ \ref{Krho}, where a comparison of the values of $K_\rho$ obtained 
for a confined system with $\tilde \rho = 1$ and those obtained by Bethe 
ansatz for $\langle n \rangle = 0.7$ is shown.\cite{giamarchi04}
 
We are now in a position to determine the conditions to attain a SS. 
This is achieved in the periodic case for $n=1$, i.e., $2 k_F a = \pi$. 
We thus increase $\tilde \rho$ until the maximum of $N(k)$ in the 
confined system appears at $k=\pi/a$. 
\begin{figure}[ht]
\includegraphics[width=.41\textwidth]{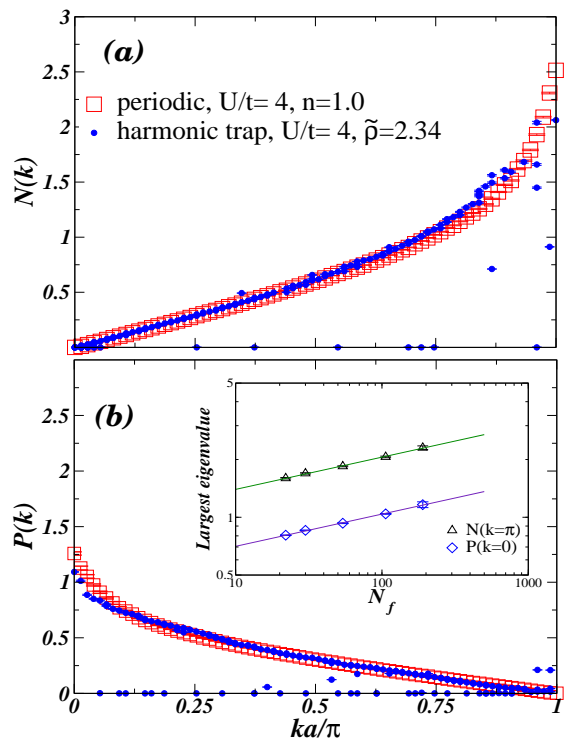}
\caption{(Color online) (a) $N(k)$ for the confined case for 
$\tilde \rho = 2.34$ ($\bullet$); other parameters as in 
Fig.\ \ref{Density}. Superimposed are the results for a periodic 
system ($\Box$) with density $n =1.0$. (b) $P(k)$ for the confined 
($\bullet$) and periodic systems ($\Box$) with the 
same parameters as in (a). Inset: largest eigenvalues for the DD 
[$N(k=\pi$)] and pairing [$P(k=0)$] vs number of fermions $N_f$
in a double-logarithmic scale.}
\label{Supersolid1}
\end{figure}
This is shown in Fig.\ \ref{Supersolid1}, where the mapping described 
above was made for both the DD [$N(k)$] and pairing [$P(k)$] eigenvalues. 
Both quantities of the confined system follow closely the behavior in the 
periodic system. As in the case of $\tilde \rho = 1$, the power-law behavior 
of the $2k_F$ oscillations in $N(k)$ for the trapped system is determined 
by the same $K_\rho$ as in the periodic case. Moreover, in the periodic case, 
due to the above mentioned SU(2) symmetry, the relationship Eq.\ (\ref{NandP}) 
holds between the correlation functions. This means in a periodic system 
that $P(k) = N(\pi/a-k)/2$, a relation that is obeyed in 
Fig.\ \ref{Supersolid1} ($\Box$). Remarkably, and to a very good 
approximation, this symmetry is seen to be present also in the confined 
system ($\bullet$). Hence, in spite of the explicit breaking of the SU(2) 
symmetry by the confining potential, this symmetry is recovered in the 
correlations, which display quasi-long-range order. This fact is reminiscent 
of the universality found for hard-core bosons confined on a 1D lattice, 
where the power-law decay of the one-particle density matrix remains 
unaffected by the presence of a confining potential.\cite{rigol04b} In fact, 
in the limit $U/t \rightarrow \infty$ of Eq.~(1), the resulting on-site 
pairs reduce to hard-core bosons. Here, we find that, due to the 
long-ranged fluctuations, once the appropriate characteristic 
density is found, the SU(2) symmetry is restored with high accuracy
for the considered values of $U$.

The inset of Fig.\ \ref{Supersolid1} shows the scaling behavior of the 
largest eigenvalues $N(k=\pi)$ and $P(k=0)$ for the harmonically confined 
system as a function of the number of particles for $\tilde \rho = 2.34$ 
and $U/t=4$. Both indeed diverge with the same power law, and, as discussed 
above, differ in magnitude within statistical errors by a factor of 2, 
as expected when the SU(2) symmetry is present. Hence, we show explicitly 
that quasi-long-range order is present in both the charge density and 
pairing channel with a unique exponent, giving rise to a SS state even 
in the trapped system.

In summary, on the basis of QMC simulations (i) we have shown that 
quasi-long-range order in a confined system of attractive fermions is characterized by the single exponent $K_\rho$ of the related periodic 
system, and given a prescription to relate both systems; (ii) we have 
also shown that this identification leads to systems where diagonal 
and off-diagonal quasi-long-range order arise with the same exponent, 
such that they correspond to a supersolid in one dimension. These
results clearly show that a SS state is attainable experimentally with 
fermionic quantum gases with two degenerate hyperfine states confined in 
one-dimensional tubes with an axial optical lattice, a configuration that 
was already realized with fermions with repulsive 
interactions.\cite{stoeferle04}

{\it Note Added}: After this work was completed and submitted, we became 
aware of a related Letter by G. Xianlong {\it et al.}\cite{xianlong07}

We acknowledge financial support by the SFB/TR 21, NSF Grants No. 
DMR-0240918, and No. DMR-0312261, and allocation of computer time at the 
HRL-Stuttgart and the NIC-J\"ulich. We thank R.\ Walser for interesting 
discussions.


\begin{thebibliography}{10}

\bibitem{kim04a}
E. Kim and M.~H.~W. Chan, 
Nature (London) {\bf 427}, 225 (2004); 
Science {\bf 305}, 1941 (2004).

\bibitem{andreev69}
A.~F. Andreev and I.~M. Lifshitz, 
Sov. Phys. JETP {\bf 29}, 1107 (1969);
G.~V. Chester, 
Phys. Rev. A {\bf 2}, 256 (1970);
A.~J. Leggett, 
Phys. Rev. Lett. {\bf 25}, 1543 (1970).

\bibitem{penrose56}
O. Penrose and L. Onsager, 
Phys. Rev. {\bf 104}, 576 (1956).

\bibitem{ceperley04}
D.~M. Ceperley and B. Bernu, 
Phys. Rev. Lett. {\bf 93}, 155303 (2004);
B.~K. Clark and D.~M. Ceperley, 
{\it ibid.} {\bf 96}, 105302 (2006).

\bibitem{prokofev05}
N.~V. Prokof'ev and B.~V. Svistunov, 
Phys. Rev. Lett. {\bf 94}, 155302 (2005).

\bibitem{burovski05}
E. Burovski {\it et~al.}, 
Phys. Rev. Lett. {\bf 94}, 165301 (2005);
S.~A. Khairallah and D.~M. Ceperley, 
{\it ibid.} {\bf 95}, 185301 (2005).

\bibitem{boninsegni06}
M. Boninsegni, N. Prokof'ev, and B. Svistunov, 
Phys. Rev. Lett. {\bf 96}, 105301 (2006).

\bibitem{kim06}
E. Kim and M.~H.~W. Chan, 
Phys. Rev. Lett. {\bf 97}, 115302 (2006) 

\bibitem{goral02}
K. G\'oral, L. Santos, and M. Lewenstein, 
Phys. Rev. Lett. {\bf 88}, 170406 (2002).

\bibitem{buechler03}
H.~P. B\"uchler and G. Blatter, 
Phys. Rev. Lett. {\bf 91}, 130404 (2003).

\bibitem{mathey06}
L. Mathey, cond-mat/0602616.

\bibitem{frustbose}
G. Murthy, D. Arovas and A. Auerbach, 
Phys. Rev. B {\bf 55}, 3104 (1997);
S. Wessel and M. Troyer, 
Phys. Rev. Lett. {\bf 95}, 127205 (2005);
R. G. Melko {\it et al.}, 
{\it ibid.} {\bf 95}, 127207 (2005);
D. Heidarian and K. Damle, 
{\it ibid.} {\bf 95}, 127206 (2005).

\bibitem{sengupta05}
P. Sengupta {\it et~al.}, 
Phys. Rev. Lett. {\bf 94}, 207202 (2005).

\bibitem{scarola05}
V.~W. Scarola and S. DasSarma, 
Phys. Rev. Lett. {\bf 95}, 033003 (2005).

\bibitem{batrouni02}
G. Batrouni {\it et~al.}, 
Phys. Rev. Lett. {\bf 89}, 117203 (2002).

\bibitem{rigol03} 
M. Rigol {\it et~al.}, 
Phys. Rev. Lett. {\bf 91}, 130403 (2003).

\bibitem{scarola06}
V.~W. Scarola, E. Demler, and S. DasSarma, 
Phys. Rev. A {\bf 73}, 051601(R) (2006).

\bibitem{micnas90}
R. Micnas, J. Ranninger, and S. Robaszkiewicz, 
Rev. Mod. Phys. {\bf 62}, 113 (1990).

\bibitem{regal04}
C.~A. Regal, M. Greiner, and D.~S. Jin, 
Phys. Rev. Lett. {\bf 92}, 040403 (2004).

\bibitem{bartenstein04}
M. Bartenstein {\it et~al.}, 
Phys. Rev. Lett. {\bf 92}, 120401 (2004).

\bibitem{zwierlein04}
M.~W. Zwierlein {\it et~al.}, 
Phys. Rev. Lett. {\bf 92}, 120403 (2004).

\bibitem{chin06}
J.~K. Chin {\it et~al.}, 
Nature (London) {\bf 443}, 961 (2006).

\bibitem{giamarchi04}
T. Giamarchi, 
{\em Quantum Physics in One Dimension} (Clarendon Press, Oxford, 2004).

\bibitem{sugiyama86}
G. Sugiyama and S.~E. Koonin, 
Ann. Phys. (N.Y.) {\bf 168}, 1 (1986).

\bibitem{sorella89}
S. Sorella {\it et~al.}, 
Europhys. Lett. {\bf 8}, 663 (1989).

\bibitem{loh92}
E.~Y. Loh and J.~E. Gubernatis, 
in {\em Modern Problems of Condensed Matter Physics}, 
edited by W. Hanke and Y. Kopaev (North-Holland, Amsterdam, 1992).

\bibitem{butts97}
D. A. Butts and D. S. Rokhsar, 
Phys. Rev. A {\bf 55}, 4346 (1997).

\bibitem{rigol04a}
M. Rigol and A. Muramatsu, 
Phys. Rev. A {\bf 69}, 053612 (2004);
Opt. Commun. {\bf 243}, 33 (2004).

\bibitem{greiner05}
M. Greiner {\it et~al.}, 
Phys. Rev. Lett. {\bf 94}, 110401 (2005).

\bibitem{rigol04b}
M. Rigol and A. Muramatsu, 
Phys. Rev. A {\bf 70}, 031603(R) (2004); {\bf 72}, 013604 (2005).

\bibitem{stoeferle04}
T. St\"oferle {\it et al}., 
Phys. Rev. Lett. {\bf 92}, 130403 (2004).

\bibitem{xianlong07} 
G. Xianlong {\it et al.},
Phys. Rev. Lett. {\bf 98}, 030404 (2007).

\end{thebibliography}
\end{document}